\documentclass[twocolumn,numbering,showpacs,superscriptaddress]{revtex4-1}

\usepackage{graphicx,color}
\usepackage{latexsym}
\usepackage{amsmath,amssymb}        
\usepackage[draft=false]{hyperref}
\usepackage{mathrsfs}
\usepackage{comment}
\usepackage{soul}
\usepackage{mathrsfs}
\definecolor{orange}{rgb}{1,0.5,0}
\DeclareSymbolFontAlphabet{\mathrsfs}{rsfs}
\DeclareMathAlphabet{\mathcal}{OMS}{cmsy}{m}{n}

\usepackage{ulem}

\begin{document}

% -----> TITLE    <-----
%\submitjournal{JCAP}

%\shorttitle{CMaDE}
%\shortauthors{Matos & L-Parrilla}

%begin{document}

\title{The graviton Compton mass as Dark Energy}

%\correspondingauthor{T. Matos}
%\email{tonatiuh.matos@cinvestav.mx}

\author{Tonatiuh Matos}
\affiliation{Departamento de F\'{i}sica, Centro de Investigaci\'on y de Estudios Avanzados del IPN, A.P. 14-740, 07000 CDMX, M\'exico.}

\author{Laura L-Parrilla}
\affiliation{Instituto de Ciencias Nucleares, Universidad Nacional
  Aut\'onoma de M\'exico, Circuito Exterior C.U., A.P. 70-543,
  M\'exico D.F. 04510, M\'exico}

\begin{abstract}
One of the greatest challenges of science is to understand the current accelerated expansion of the Universe. In this work, we show that by considering the quantum nature of the gravitational field, its wavelength can be associated with an effective Compton mass. We propose that this mass can be interpreted as dark energy, with a 
Compton wavelength given by the size of the observable Universe, implying that the dark energy varies depending on this size.  If we do so, we find that: 1.- Even without any free constant for dark energy, the evolution of the Hubble parameter is exactly the same as for the LCDM model, so we expect this model to have the same predictions as LCDM. 2.- The density rate of the dark energy is $\Omega_\Lambda=0.69$ which is a very similar value as the one found by the Planck satellite $\Omega_\Lambda=0.684$. 3.- The dark energy has this value because it corresponds to the actual size of the radius of the Universe, thus the coincidence problem has a very natural explanation. 4.- It is possible to find also a natural explanation to why observations inferred from the local distance ladder find the value $H_0=73$ km/s/Mpc for the Hubble constant. We show that if we take the variability of the dark energy into account, they should measure $H_0=67.3$ km/s/Mpc as well. 5.- In this model the inflationary period contains a natural successful graceful exit.
\end{abstract}

\pacs{Cosmological Constant -- Hubble Parameter-- Compton Mass}

\maketitle

\section{Introduction} \label{sec:intro}

Nowadays, the mystery of the nature of so-called dark energy is one of the greatest challenges of science. It consists in to find out the reason why the universe is expanding with some acceleration, which means, understanding the current accelerated expansion of the Universe. \cite{Tawfik:2019dda}. There are many hypotheses, from modifications of the Einstein equations to the proposal of exotic forms of matter, but the cosmological constant continues to be one of the most accepted candidates \cite{Wondrak:2017eao}. Nevertheless, all these hypotheses people are working with today have some problems; most of them contain some discrepancy between the values obtained from cosmology and the corresponding values obtained using current observations. One of the most accepted hypotheses is the cosmological constant, related somehow with the vacuum expectation value \cite{Mortonson:2013zfa}. However, there seems to be no way to get the value of this constant using simple arguments. Today, the most accepted model is a cosmological constant to explain the accelerated expansion of the universe together with a hypothetical particle that behaves as dust, modeling the dark matter. Both hypotheses together are the so-called Lambda Cold Dark Matter (LCDM) model. On the other hand, in recent times, using this LCDM model, there is tension between observations of the Planck satellite obtained using the CMB fluctuations and the value of the Hubble parameter $H_0$ measured using other methods. While the Planck satellite gives the value $H_0=0.684$ km/s/Mpc \cite{Akrami:2018vks}, the observations inferred from the local distance ladder give the value $H_0=0.73$ km/s/Mpc \cite{Riess:2016jrr}. Nowadays, there is a consensus that this discrepancy could be because we are forgetting some important physics in the analysis of the problem.

In this work, we will give a possible solution for the two last problems using very simple arguments for the gravitational interaction. In order to do so, we start from the assumption that the gravitational interaction is quantum mechanical, let us remind the reader about one of the most important features of quantum particles. In the 1920s, Arthur Compton discovered in a scattering experiment between light and electrons. That particles contain an effective wavelength given by $\lambda=h/m c$, where $\lambda$ is the associated wavelength of a particle of mass $m$. Here $h$ is the Planck constant and $c$ the speed of light \cite{Weinberg}. On the other hand, because of the wave-particle duality of quantum objects, one can associate an effective mass $m_\gamma$ to a wave with frequency $\nu$ and energy $E=h\nu=m_\gamma c^2$. Therefore, another well-known way of interpreting Compton scattering is to say that, at this energy, the behavior of the photon as a particle with mass $m_\gamma = h\nu/c^2$ dominates. The rest mass of the photon is still zero, but it could be interpreted as a particle with this effective mass.
Strictly speaking, this implies that if the mass of a particle is zero, the corresponding Compton wavelength should be infinite. However, the Universe is finite, and therefore the wavelength of any particle must be finite as well. In this work, this is the fact that we want to use and show that this can imply the existence of effective dark energy. We call it the Compton Mass Dark Energy (CMaDE) in order to distinguish it from other proposals. Here, it is important to note that the gravitational interaction has no real mass; the rest mass of the gravitational interaction remains zero. The mass $m$ is associated with it because the gravitational interaction can be interpreted as a particle, the graviton, but as a wave, the gravitational interaction has a maximum wavelength limited by the size of the observable Universe, and with this wavelength, we can associate an effective mass $m$ using Compton's formula. As we shall see, this mass is so small that the gravitational interaction effective mass, or in other words, the behavior of the graviton as a particle, is perceptible only at cosmological scales.

\section{The main idea}

The idea in this work has two hypotheses. 

1.- Since we are assuming that gravitation is a quantum mechanical interaction; it must meet all the characteristics of a quantum particle, so it has a Compton effective mass. 

2.- Since there is strong evidence that our universe has a beginning and is limited, the wavelength $\lambda$ of the gravitational interaction is limited by the size of the observable Universe; therefore, the wavelength of the interaction particle is the path that the gravitational interaction has traveled through the Universe. 

Therefore, the effective mass of the gravitational interaction can be determined by the Compton formulas $m=h/\lambda c$, but now it is applied to the gravitational interaction. This implies that the gravitational interaction does indeed have an effective mass; so it must follow an equation similar to Proca. To see this, we write the massless and mass field equations in the linearized regime
The gravitational field $g_{\mu\nu}$ is linearized as $g_{\mu\nu}=\eta_{\mu\nu}+h_{\mu\nu}$, where $\eta_{\mu\nu}$ is the Minkowski metric and $h_{\mu\nu}$ is the weak field metric such that $|h_{\mu\nu}|<<1$. In terms of $h_{\mu\nu}$, the vacuum field Einstein equations can be written as \cite{Weinberg2}
\begin{equation}
   2 R_{\mu\nu}\sim \Box{h_{\mu\nu}}=0,
\end{equation}
being $R_{\mu\nu}$ the Ricci tensor and  $\Box$ the d'Alambert operator. This is interpreted as the gravitational waves or as the massless graviton equation. Nevertheless, the CMaDE gravitational interaction has an effective mass; therefore, the corresponding Einstein equation is now 
\begin{equation}\label{eq:KK}
    \Box{g_{\mu\nu}}-\frac{m^2c^2}{\hbar^2}g_{\mu\nu}=0,
\end{equation}
where $m$ is the effective mass associated to $h_{\mu\nu}$ using the Compton's prescription. However, $m$ is no longer a constant because it depends on its wavelength that is determined by the size of the observable Universe, which is in expansion. Nevertheless, as we will see later, $m$ varies very slowly after inflation; therefore, we can neglect a dynamical equation for $m$ as a very good approximation. 
 If we compare (\ref{eq:KK}) with the Einstein equations in vacuum, $G_{\mu\nu}+\Lambda g_{\mu\nu}=0$, we find that they are related as \cite{Weinberg2} 
 \begin{equation}\label{eq:Lam1}
     2\Lambda=\frac{m^2c^2}{\hbar^2}.
 \end{equation}
 Thus, we may identify the CMaDE $\Lambda$ with the effective mass of the gravitational interactions. In other words, we may identify the CMaDE as the energy of a vibration from the gravitational interaction because it is confined within the Universe horizon. The vibration frequency $\nu$ of the gravitational interaction is directly related to the CMaDE by $\Lambda=2\pi^2 \nu^2/c^2$. Thus, we can interpret this vibrations as the cause of a pressure that expands the Universe accelerated.
 As we shall see, this is enough to explain the accelerated expansion of the Universe as we see it.
 Now we use in (\ref{eq:Lam1}) the relationship of the Compton mass with the wavelength of the gravitational interaction to obtain that
\begin{equation}\label{eq:Lam}
     \Lambda=\frac{2\pi^2}{\lambda^2}.
 \end{equation}
The wavelength $\lambda$ is limited by the size of the observable Universe. If the gravitational interaction travels a distance $\lambda|_{today}$ during its life, the wavelength will be $\lambda|_{today}=(c/H_0)\, R_H$ long, where the unitsless quantity $R_H$ is defined as
\begin{equation}\label{eq:RH}
 R_H=H_0\int_0^{today}\frac{dt}{a}=\int_{-\infty}^0\frac{H_0}{H}e^{-N}dN,
 \end{equation}
given in terms of the e-folding parameter $N=\ln(a)$ and the Hubble parameter $H=\dot N$, being $a$ the scale factor of the Universe. 
\begin{figure}
\centering
\includegraphics[width=0.44\textwidth]{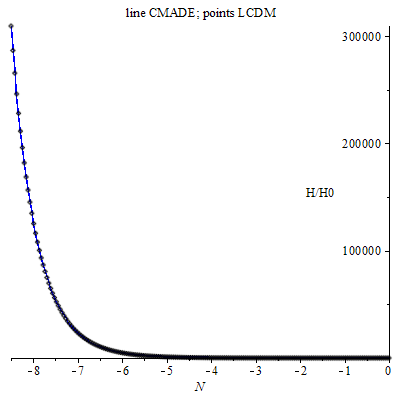}
\caption{The evolution of the Hubble parameter using the CMaDE (solid line) and the LCDM model (point line). We used the Planck values $\Omega_m=0.315$, $\Omega_r=10^{-4}$, $\Omega_\Lambda=0.684$ and $H_0=67.3$ km/s/Mpc in both plots.}
\label{fig:HNumerico}
\end{figure}
Here it is important to note that given (\ref{eq:Lam}) with (\ref{eq:RH}) for the function $\Lambda$ implies that the CMaDE model does not have free constants to fit the observations. Also, note that, because $\Lambda$ is not a constant, the Bianchi identities have an extra term 
\begin{equation}\label{eq:Ldot}
    \dot\Lambda=H\frac{d\Lambda}{dN}=-\frac{4\pi^2c}{\lambda^3}\exp(-N).
\end{equation}
Nevertheless, this term is important only before inflation. To see this, we know that $z\sim 10^{26}$ is the corresponding redshift for inflation, that implies $N\sim -60$. Just before inflation $\lambda=\lambda_0$ is small, and the exponential factor is big. Therefore, before inflation, the term (\ref{eq:Ldot}) is an extra term for the Bianchi identities. Nevertheless, after inflation $\lambda$ grows up, let say, $e^{60}$ times, thus $\lambda\sim \lambda_0e^{60}$ is huge, it grows enormously, and term (\ref{eq:Ldot}) goes very fast to zero. Thus, after inflation, the Bianchi identities are exactly fulfilled.

Observations show that the universe is almost flat, so in this work we will assume a flat space-time as a first approximation. Now we find the corresponding Freedman equation for the CMaDE model. It reads
\begin{equation}\label{eq:Freedman}
     H^2=\frac{\kappa^2}{3}\left(\rho_m+\rho_r+\rho_\Lambda\right),
 \end{equation}
where $\kappa^2=8\pi G/c^4$ is the Einstein's constant, $\rho_m$ stands for the matter density of the universe, $\rho_r$ for the radiation and $\rho_\Lambda=\Lambda/c^2\kappa^2$ for the dark energy density. If we substitute the equations (\ref{eq:Lam}) and (\ref{eq:RH}) into the derivative with respect to $N$ of the equation (\ref{eq:Freedman}), and solve the first derivative of $H$ with respect to $N$, after some manipulations, we obtain that
\begin{eqnarray}\label{eq:Freedman2}
      \frac{HH'}{H_0^2}+\frac{3}{2}\Omega_m e^{-3N}+2\Omega_r e^{-4N}&-&\nonumber\\\sqrt{\frac{3}{2}}\frac{H_0e^{-N}}{H\pi}\left(\frac{H^2}{H_0^2}-\Omega_m e^{-3N}-\Omega_r e^{-4N}\right)^{\frac{3}{2}}&=&0,
\end{eqnarray}
where a prime stands for the derivative with respect to the N-folding parameter $N$ and $\Omega_x=\rho_{0x}/\rho_{crit}$, being $\rho_{crit}=3H_0^2/\kappa^2$ the actual critical density of the Universe. Then, equation (\ref{eq:Freedman2}) is the corresponding Freedman equation for the CMaDE. The main result of this work is that this equation could very well explain the accelerated expansion of the Universe without any extra free constant and any fine-tuning.

\section{Some results}

It is possible to solve equation (\ref{eq:Freedman2}) numerically. The result is shown in fig.\ref{fig:HNumerico} where
we compare the numerical solution of (\ref{eq:Freedman2}) with the evolution of $H$ using the LCDM model in terms of the e-folding parameter, $H_{LCDM}=H_0\sqrt{\Omega_m e^{-3N}+\Omega_r e^{-4N}+\Omega_\Lambda}$, being $\Omega_m$, $\Omega_r$ and $\Omega_\Lambda$ the density rates of the Universe for the matter, radiation and dark energy, respectively \cite{Kisslinger:2019ysx}. Note that the Hubble parameter for CMaDE evolves exactly like the LCDM model; this implies that they have the same predictions.
As in the LCDM case, the CMaDE density remains subdominant all the time. Clearly, both functions are really very similar.
%Therefore, in what follows, in order to see more clearly the physical content of this hypothesis, we use the $H_{LCDM}$ as a good approximation. We do so for two reasons: first, because the observation values with which we want to compare are obtained using the LCDM model, so we must also use it; and second, because to see the physical content we do not need to get lost in the calculations, fig.\ref{fig:HNumerico} shows that both $H$'s are very similar. For example, we can use the  $H_{LCDM}$ to find the value of $R_H$.
Using the numerical integration of (\ref{eq:Freedman2}), we can integrate (\ref{eq:RH}). Observe that this integral does not have any integration constant; we find that  $R_H=3.087$.  If we put this value of $R_H$ in (\ref{eq:Lam}), we get that
\begin{equation}\label{eq:Lam2}
     \Lambda=2\left(\frac{\pi}{3.087}\right)^2\frac{H_0^2}{c^2}=\frac{3H_0^2}{c^2}\Omega_\Lambda,
 \end{equation}
 where we find that $\Omega_\Lambda=0.69$. 
Remarkably, that this theoretical value is in very good agreement with the observed value of the Planck satellite $\Omega_\Lambda=0.684$. Note that this value of $\Omega_\Lambda$ strongly depends on the size of the wavelength (\ref{eq:RH}). On the other hand, the extreme similarity of the Hubble parameter $H$ in the CMaDE and LCDM models, guarantees that the predictions of both models are the same. Furthermore, this fact is supported by simulations given in \cite{Vazquez:2012ag}, where comparisons with observations with a very similar models were performed.

This result also gives an explanation of the coincidence problem, because the value of the CMaDE now is determined by the size of the Universe horizon, which determines the value of the size $\lambda$ of the wavelength.

Particularly, during the matter dominated epoch $H=1/t=H_0/a^{2/3}$ \cite{Weinberg2}, one finds that $R_H$ evolves as $R_H=2c\sqrt{a}/H_0$. Thus, we have that during the matter dominating epoch
\begin{equation}\label{eq:Lam3}
     \Lambda=\frac{\pi^2}{6}\frac{3H_0^2}{c^2}\frac{1}{a}.
 \end{equation}

Thus, the field equation for $\Lambda$ is just $\dot\Lambda+H\Lambda=0$. Using this approximation, it is easy to see that the Hubble parameter evolves as 
\begin{equation}\label{eq:H0}
     H=H_0\sqrt{\Omega_m e^{-3N}+\Omega_r e^{-4N}+\Omega_\Lambda e^{-N}}.
 \end{equation}
 
\begin{figure}
\centering
\includegraphics[width=0.44\textwidth]{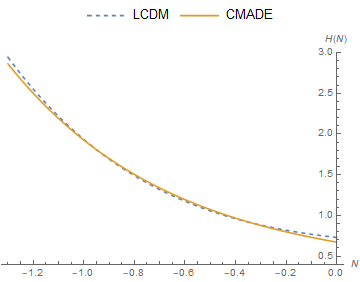}
\caption{The evolution of the Hubble parameter using $H_{LCDM}$ (dashed line) and a variable CMaDE (solid line) (\ref{eq:H0}). We observe how the same values of the Hubble parameter for $N\sim-1$ give different values for $H$ in $N=0$. In other words, if the local distance ladder measure a value of $H$ at redshift $z\sim 2$ and we do not take the variability of the CMaDE into account, we obtain a false value for $H_0$ today. We used the Planck values $\Omega_m=0.315$, $\Omega_r=10^{-4}$, $\Omega_\Lambda=0.684$ in both plots and $H_0=67.3$ km/s/Mpc for the variable $\Lambda$ (solid line) and $H_0=73$ km/s/Mpc for a constant $\Lambda$ (dashed line). Notice that both lines are very similar. This similarity goes further into the era of radiation dominance because CMaDE varies very slowly throughout this time.}
\label{fig:Hevolution}
\end{figure}

 We compare the evolution of the Hubble parameter of CMaDE (\ref{eq:H0}) with the LCDM model using a cosmological constant with the same values for the $\Omega$'s, we show this in Fig.\ref{fig:Hevolution}. Here we use in both evolutions the best values given by the Planck satellite.  We see that both evolve in a very similar manner, but given almost the same values for $N\sim -1$, just in the region where the local distance ladder observations take place. Even when the evolution of $H_{LCDM}$ reaches the value $H_0=0.73$ km/s/Mpc, while the variable CMaDE reaches $H_0=0.673$ km/s/Mpc, they have the same values in a large region near $N\sim -1$. Thus, we conjecture that if the observations using the local distance ladder considering the small variation of the CMaDE, they should obtain the same values as the corresponding ones measured by the Planck satellite. 

%\section*{Results}

In conclusion, if we consider the quantum nature of the gravitational field, this may imply that it has a quantum Compton effective mass that we may feel as a variable CMaDE. Taking this into account, we could explain the actual value of the cosmological constant, the coincidence problem, and we could give a natural explanation to the tension  for the value of $H_0$ obtained by the Planck satellite and the one inferred from the local distance ladder observations. 
 
It remains to check whether the evolution of the fluctuations of the universe with CMaDE evolve as those observed. We can expect it to be so; because the changes for LCDM in the matter-dominated epoch are so small that the differences must also be small (see for example \cite{Vazquez:2012ag}). In the radiation dominated epoch the variable CMaDE evolves as $\Lambda\sim 1/a^2$, but in that period the main observational constraint is the measurements of the Big Bang Nucleosynthesis which is essentially determined by radiation content of the Universe, which is not altered here. Thus, we expect that all the present cosmological observations of the Universe are in good agreement with our hypotheses. The results presented here may be a simple coincidence of numbers, but we agree that all fundamental interactions in nature are in fact quantum mechanical, including gravitational, and the results presented here all stem from this fact. The main result of this work there may be no exotic matter responsible for the accelerated expansion of the Universe. The present work shows that this expansion may be a simple consequence of the quantum nature of the gravitational interaction.

\section{Inflationary epoch}
 
 It remains to study the behavior of this hypothesis at the origin of the universe, where the graviton wavelength is small and, therefore, the effective mass of the graviton is large. Unfortunately, in this region, the quantum characteristics of the gravitational field are important and we cannot decide what happened so far, because we do not have a theory of quantum gravity. However, we can speculate some features of that origin. Just after the Planck time, we can suppose that there exists an inflaton field in the standard way. Besides the $\Lambda$ function, we add the inflaton field $\phi=\phi(t)$ to the Einstein equations. However, this scalar field  is here non conserved, such that the Bianchi identities are now
 \begin{equation}\label{eq:phi}
     \kappa^2\dot\phi\left(\Box\phi-\frac{dV}{d\phi}\right)=\dot\Lambda
 \end{equation}
 being $V$ the inflaton potential. We can rewrite the function $\Lambda=\Lambda(t)$ as a function of $\Lambda=\Lambda(\phi)$, such that equation (\ref{eq:phi}) can be rewritten as
 \begin{equation}\label{eq:phi2}
     \Box\phi-\frac{dV}{d\phi}-\frac{1}{\kappa^2}\frac{d\Lambda}{d\phi}=0
 \end{equation}
 This implies that the inflaton potential is now endowed with the cosmological function $V\longrightarrow V+1/\kappa^2\Lambda$. Following the same procedure as for equation (\ref{eq:Freedman2}), the Friedman equation transforms into
 \begin{eqnarray}\label{eq:FreedmanInf}
      \frac{HH'}{H_0^2}+\frac{1}{2}\Omega'_\phi+ 
      \sqrt{\frac{3}{2}}\frac{H_0e^{-N}}{H\pi}\left(\frac{H^2}{H_0^2}-\Omega_\phi\right)^{\frac{3}{2}}=0,%\nonumber\\ 
\end{eqnarray}
 where $\Omega_\phi=\rho_\phi/\rho_{crit}$ being $\rho_\phi=1/2\dot\phi^2+V$ the scalar field density. After the inflationary extreme expansion, the universe grows up an enormous amount, and the wavelength of the gravitational interaction (\ref{eq:Lam}) grows hundreds of orders of magnitude, causing that the function $\Lambda$ decays very fast, becoming very, very small. After that, the function $H$ decays to a very small value. This stops inflation naturally. In the meanwhile, the quarks and leptons form and build radiation dominated Universe; thus, the Hubble parameter changes its behavior to $H\sim 1/t\sim 1/a^2$. The $\Lambda$ parameter then behaves as $\Lambda\sim 1/a^2$, and it continues the history of the Universe as LCDM.
 
However, we think that this hypothesis should be further studied, but it certainly opens a new window of research.
 
 \acknowledgments

We thank Jorge Cervantes for very helpful discussions and Alberto V\'azquez for their remarks after a deep reading of the paper. This work was partially supported by CONACyT M\'exico under grants  A1-S-8742, 304001, 376127;
Xiuhcoatl and Abacus clusters at Cinvestav, IPN;
I0101/131/07 C-234/07 of the Instituto
Avanzado de Cosmolog\'ia (IAC) collaboration (http://www.iac.edu.mx/). This research received support by Conacyt through the Fondo Sectorial de Investigaci\'on para la Educaci\'on, grant No. 240512.


\bibliography{Compton}{}

\end{document}